\documentclass[aps,numerical,graphicx,showpacs,article,prl]{revtex4-1}
\usepackage{amsmath}
\usepackage{graphicx}

\begin{document}

\def\Journal#1#2#3#4{{#1 }{\bf #2, }{ #3 }{ (#4)}}

\def\BiJ{ Biophys. J.}
\def\Bios{ Biosensors and Bioelectronics}
\def\LNC{ Lett. Nuovo Cimento}
\def\JCP{ J. Chem. Phys.}
\def\JAP{ J. Appl. Phys.}
\def\JMB{ J. Mol. Biol.}
\def\JPC{ J. Phys: Condens. Matter}
\def\CMP{ Comm. Math. Phys.}
\def\LMP{ Lett. Math. Phys.}
\def\NLE{{ Nature Lett.}}
\def\NPB{{ Nucl. Phys.} B}
\def\PLA{{ Phys. Lett.}  A}
\def\PLB{{ Phys. Lett.}  B}
\def\PNAS{Proc. Natl. Am. Soc.}
\def\PRL{ Phys. Rev. Lett.}
\def\PRA{{ Phys. Rev.} A}
\def\PRE{{ Phys. Rev.} E}
\def\PRB{{ Phys. Rev.} B}
\def\PD{{ Physica} D}
\def\ZPC{{ Z. Phys.} C}
\def\RMP{ Rev. Mod. Phys.}
\def\EPJD{{ Eur. Phys. J.} D}
\def\SAB{ Sens. Act. B}
\title{
From conductance viewed as transmission to resistance viewed as reflection.
An extension of Landauer quantum paradigm to the classical case
at finite temperature
}
\author{Lino Reggiani}
\email{lino.reggiani@unisalento.it}
\affiliation{Dipartimento di Matematica e Fisica, ``Ennio de Giorgi'',
Universit\`a del Salento, via Monteroni, I-73100 Lecce, Italy}
\affiliation{CNISM,  Via della Vasca Navale, 84 - 00146 Roma, Italy}

\author{Eleonora Alfinito}
\affiliation{Dipartimento di Matematica e Fisica, ``Ennio de Giorgi'',
Universit\`a del Salento, via Monteroni, I-73100 Lecce, Italy}
\author{Federico Intini}
\affiliation {Department of Sciences and Methods for Engineering \\
Via Amendola 2, Pad. Morselli - 42122 Reggio Emilia, Italy
}
\date{\today}
\begin{abstract}
In this paper we present an extension of Landauer paradigm, conductance is transmission,  to the case of macroscopic classical conductors making use of a  description of conductance and resistance based on the application of the fluctuation dissipation (FD) theorem.
The main result is summarized in the expressions below for conductance 
$G$ and resistance $R$ at thermodynamic equilibrium, with the usual meaning of symbols. 
$G$ is given in terms of the variance of total carrier number fluctuations between two ideal transparent contacts in an open system described by  a grand canonical ensemble as
%
$$
G
=\frac{e^2 \overline{v_x'^2} \tau}{L^2 K_BT} \overline{\delta N^2} 
$$
%
By  contrast $R$ is given in terms of the variance of carrier drift-velocity fluctuations  due to the instantaneous carrier specular  reflection at the internal contact interfaces of a closed system described by a  canonical ensemble as 
$$
R
= \frac{(m L)^2}{e^2 K_BT \tau} \overline{\delta v_d^2}
$$
The FD approach gives evidence of the duality property of conductance related to transmission and resistance related to reflection. 
Remarkably, the expressions above are shown  to recover the quantum Landauer paradigm in the limit  of zero temperature for a one-dimensional 
conductor.

%
\end{abstract}
\maketitle 
%
\section{Introduction} 
Conductance is transmission is a famous paradigm credited to Landauer since 1957 \cite{landauer57,imry99}, when he  proposed that conductance at the nanometer scale length  in a 1D quantum conductor could be viewed as transmission. 
Then, in the presence of scattering conductance is quantized into the sum of an integer number of fundamental conductance unit $G_0$ as
\begin{equation}
G =G_0 \Sigma_i \Gamma_i
\end{equation}
with the integer $i$ labeling the number of transverse mode involved, 
$\Gamma_i$ is the respective transmission probability, and
\begin{equation} 
G_0=2\frac{e^2}{h}
\end{equation}
with $e$ the unit of electric charge and $h$ the Planck constant.
\par
For the simple case of a single transport channel (i=1), balistic transport,
i.e. $ \Gamma_1 = 1$, G takes the  value of the fundamental unit of conductance  $ G_0=8.12 \times 10^{-5} 	\ \Omega^{-1}$.
We conclude that the concept that conductance can be viewed as transmission depends only from the appearance and the  values of the quantum transmission probability $\Gamma_i$.  
Furthermore, using the reciprocity property, the quantum
resistance $R=1/G$ is found to depend on the inverse of the transmission probability. 
Therefore, within the Landauer model there is only reciprocity between 
$G$ and $R$.
This is due to the theoretical framework used by Landauer that makes use of linear response theory under full degenerate conditions, 
that is at temperature $T=0$.
\section{theory} 
We consider a macroscopic homogenous sample,  
of length $L$ and cross-sectional area $A$, characterized by a measurable intrinsic conductance (resistance) following Ohm law, 
under thermal equilibrium conditions at a given  temperature $T$.
To avoid boundary effects we assume  $A \gg L^2$, then at the extremes 
we take two ideal electrical contacts as detailed later according to the  measurements conditions of constant voltage or constant current. 
Then, both $G$ and $R$ are deermined at thermal equilibrium making
use of the fluctuation-dissipation (FD) theorem (Nyquis relations) 
\cite{nyquist28,kubo66}.  
\par
Following a recent work on the reciprocity and dual properties of conductance and resistance  of an Ohmic conductor 
\cite{reggiani16}, we investigate the possibility to extend the Landauer paradigm to the classical case in the presence of a finite temperature by including also  the resistance viewed as reflection.
We would stress, that here transmission is associated  with  the total number of charge carriers transmitted through contacts and reflection with the  charge carrier drift-velocity internally reflected by contacts.
\par
\begin{figure}
 \begin{center}
 \includegraphics[width=15cm]{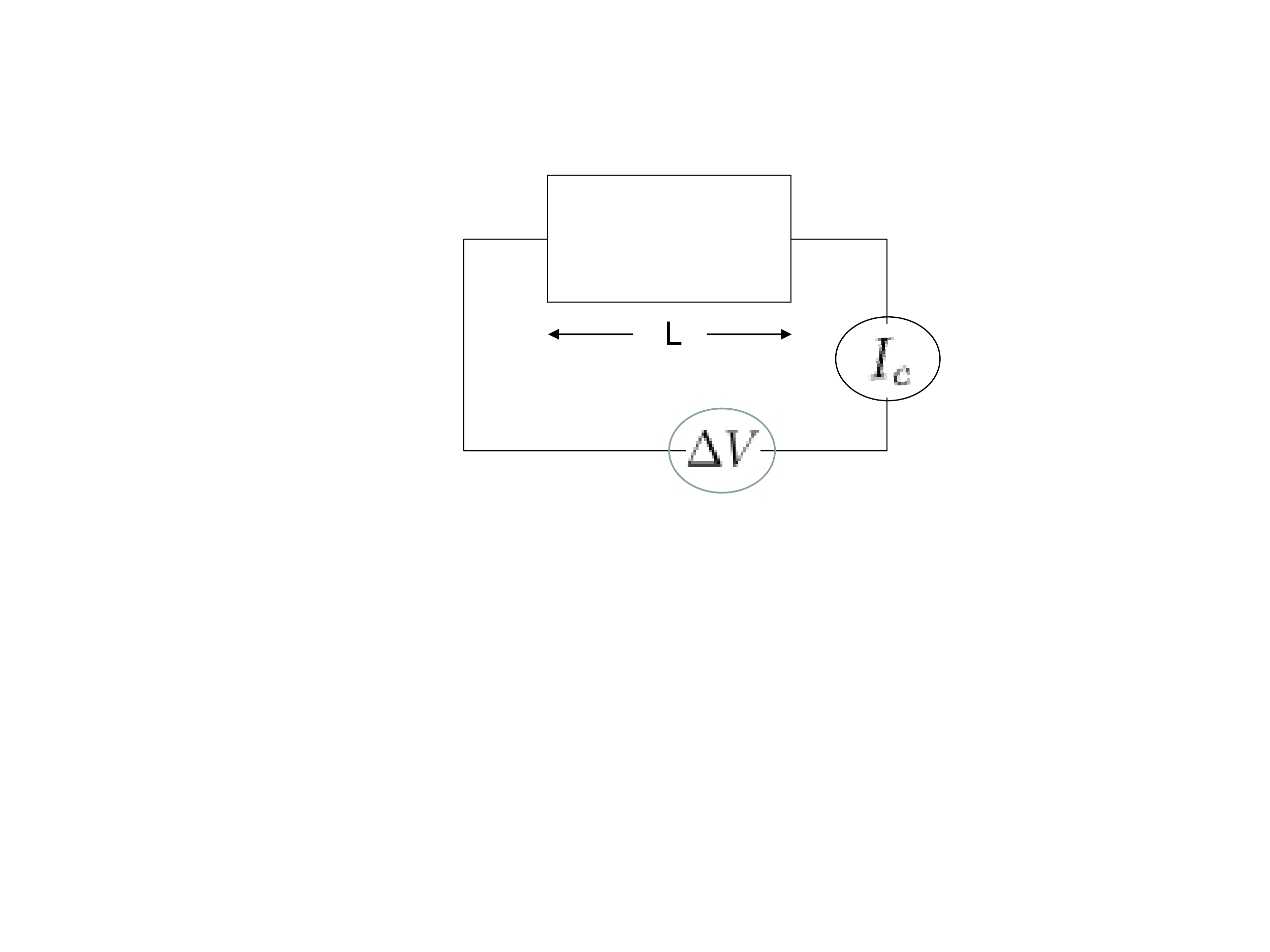}
 \end{center}
\vspace{-40mm}
 \caption{ Schematic of the circuit used to determine resistance/conductance  from current-voltage measurements of a given two terminal sample at temperature $T$.
}
\label{figiv}
\end{figure}

 
According to Ohm law, for a homogeneous conductor the reciprocity property gives:
\begin{equation}
G=\frac{1}{R} = \frac{I}{V} 
\end{equation}
with $I$ and $V$, respectively  the current and voltage drop involved in the experiment as sketched in Fig.~(\ref{figiv}).
\subsection{The classical diffusive theoretical-model}
\par
Below we briefly survey  a series of definitions of $G$ and $R$ that summarize the reciprocity i.e.  $GR=1$, according to kinetic models that are based on the following characteristics: 3D  diffusive (presence of scattering).
\par
From a diffusive (dif) kinetic model (Drude 1900), conductance and its reciprocal resistance are given by:
\begin{equation}
G^{dif} = \frac{1}{R^{dif}} = \frac{e^2 \tau \overline N} {L^2 m} 
\end{equation}
with $e$ the unit charge,  $\tau$ the scattering time, $\overline N$ the average  number of free carriers inside the sample,  and $m$ their effective mass.
\par
From the generalized Einstein relation (Einstein 1905 and Smoluchowski 1906) \cite{einstein05} 
it is:
\begin{equation}
G^{dif}=\frac{1}{R^{dif}}=\left(\frac{e}{L}\right)^2 D^{dif}_x  
\frac{\partial \overline{ N}} { \partial \mu_0} 
\end{equation}
with 
\begin{equation}
D_x^{dif}=\overline{v^{2'}_x} \tau = \frac{\overline{N} \tau}{m}  
\frac{\partial \mu_0} {\partial \overline{ N}} 
\end{equation}
the longitudinal diffusion coefficient,
$\mu_0$ the chemical potential with the differential
(with respect to carrier number) quadratic velocity
component along the x direction given by \cite{gurevich79}
\begin{equation}
\overline{v_x^{2'}} = \frac{ K_BT}{m} \frac{\overline{N}}
{\overline{\delta N^2}}
\end{equation}
\par
Finally, the kinetic lumped-model gives:
\begin{equation}
R^{dif} = \frac{\tau_d}{\cal C}
\end{equation}
with the capacitance
\begin{equation}
{\cal C}=\frac{A}{L} \epsilon_0 \epsilon_r
\end{equation}
and $\tau_d$ the dielectric relaxation time.
\par
\begin{figure}
 \begin{center}
  \includegraphics[width=15cm]{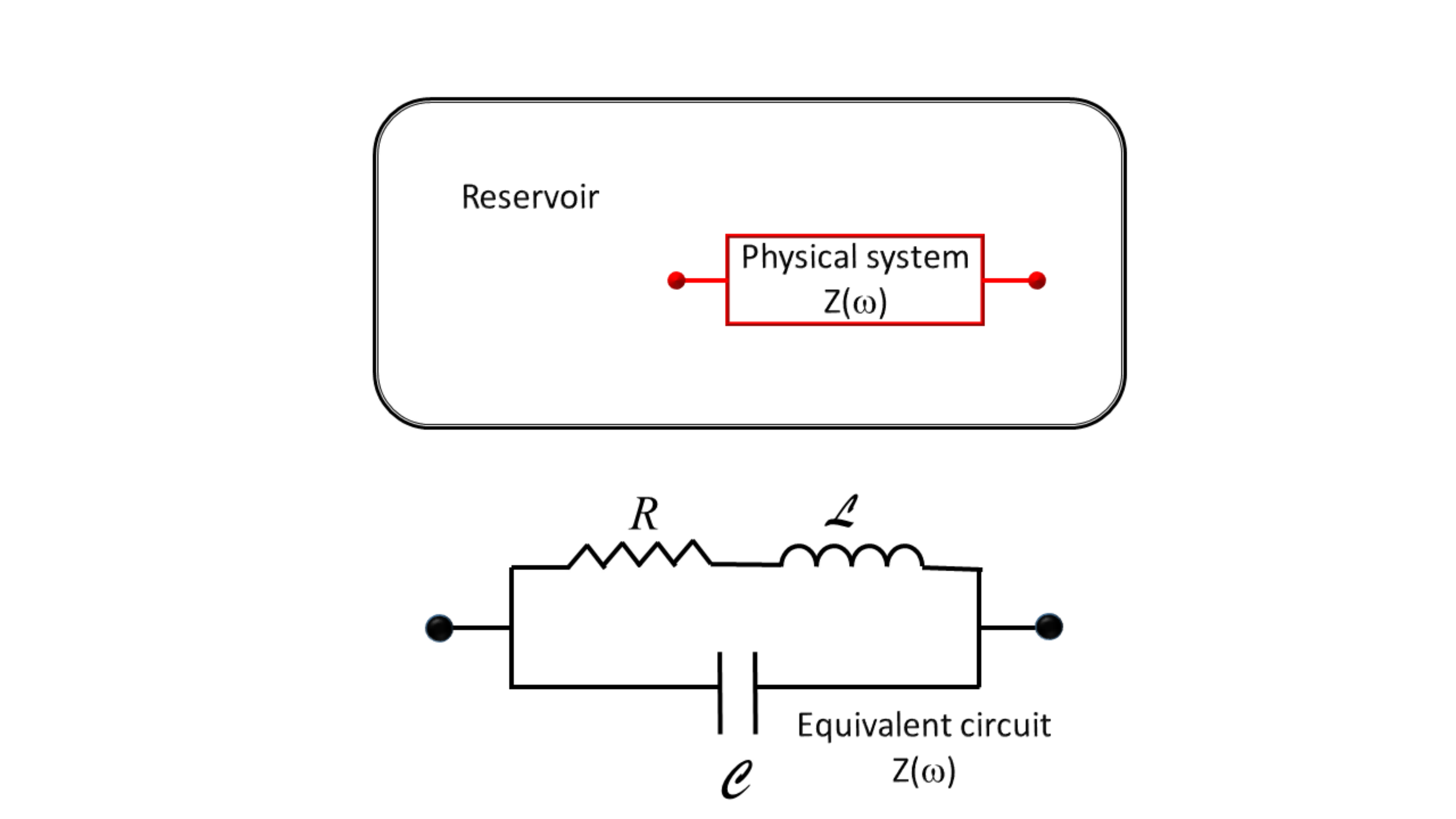}
 \end{center}
\vspace{-2mm}
 \caption{ Schematic of the equivalent circuit of the intrinsic impedance $Z(\omega)$ that consists of a resistor $R$, a  kinetic-inductance ${\cal L}$   and a parallel 
plates capacitor $\cal {C}$ filled with the homogenous medium that constitutes the resistor of given relative dielectric-constant.
The capacitance  and inductance  account
for the presence of the contacts and for the inertia of carriers, respectively.
The reservoir can be a grand-canonical or a canonical ensemble  according to the operation conditions for noise detection at constant current or constant voltage operation modes, respectively.}
\label{figec}
\end{figure}
\subsection{The classical three dimensional case in the FD model}
For the analysis of current or voltage fluctuations at a kinetic level a
correct system definition becomes of prime importance. 
On the one hand, the microscopic model for carrier transport implies a well-defined equivalent circuit. 
On the other hand, the measurement of current or voltage
fluctuations in the outside circuit is reflected in the boundary conditions for the microscopic modeling, that determine the choice of the appropriate statistical ensemble. 
Current noise is measured in the outside short-circuit,
which implies an open system where carriers may enter or leave the sample,
thus referring to the grand canonical ensemble (GCE). 
Voltage noise is measured in the outside open circuit when the carrier number in the sample is fixed, thus referring to the canonical ensemble 
(CE). 
\par
The main items of the theoretical approach are based on  previous 
papers \cite{greiner00, reggiani16,reggiani18} and are briefly recalled  in the following.
To extend the Landauer paradigm to macroscopic conductors we use the fluctuation-dissipation  (FD) theorem, i.e. a thermodynamic equilibrium approach to conductance/resistance that implies the presence of a temperature $T$ different from zero.
Accordingly, we found:
\begin{equation}
G^{FD}
=\frac{e^2 \overline{v_x'^2} \tau}{L^2 K_BT} \overline{\delta N^2} 
= \frac{e^2 \overline{N}  \Gamma } {Lm\sqrt{\overline{v_x'^2}}} 
\label{eGc}
\end{equation}
with
\begin{equation}
\Gamma=\frac{l}{L}=
\frac{\tau  \sqrt{\overline {v_x'^2 }} }{L}
\end{equation}
being $l$ the carrier mean free path and  $\Gamma$ the classical transmission probability for a carrier to cross the full sample in the presence of scattering (not to be confused with the transmission probability to cross an interface), notice that $0 <\Gamma \le 1$, and in the classical  balistic case $\Gamma =1$.
Remarkably, the first form of Eq. (\ref{eGc}) relates conductance to the variance of total carrier number fluctuations.
By contrast the second form of Eq. (\ref{eGc}) represents the 3D diffusive analog of the Landauer formula for the fundamental quantum unit of electrical conductance.
\par
\begin{figure}
 \begin{center}
  \includegraphics[width=15cm]{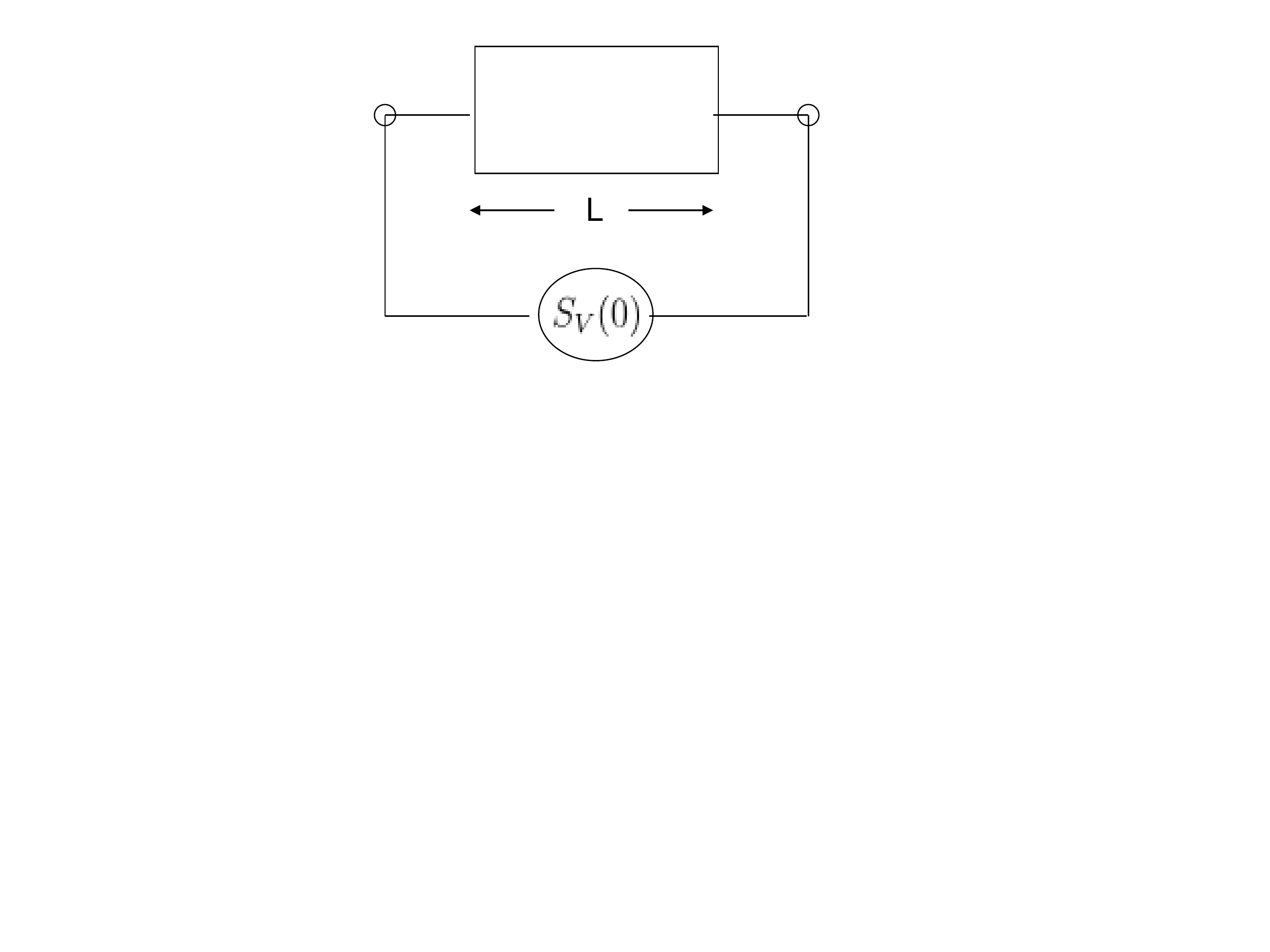}
 \end{center}
\vspace{-50mm}
 \caption{ Schematic of the circuit used to determine resistance from
voltage fluctuations due to carrier reflections at the contacts inside the sample, as measured in the external open-circuit.}
\label{figsv}
\end{figure}

Within the same approach, the definition of resistance from he FD theorem gives:
\begin{equation}
R^{FD} 
= \frac{(m L)^2}{e^2 K_BT \tau} \overline{\delta v_d^2}
= \frac {Lm\sqrt{\overline{v_x'^2}}} {e^2 \overline{N}  \Gamma }
\label{eRc}
\end{equation}
Remarkably, the first form of Eq. (\ref{eRc}) relates resistance to the variance of carrier drift-velocity fluctuations.
By contrast the second form of Eq. (\ref{eRc}) represents the 3D diffusive analog of the Landauer formula for the fundamental quantum unit of electrical resistance.
The first equation of the above definitions summarize the duality properties according to models that are based on the  microscopic sources of carrier fluctuations. 
\begin{figure}
 \begin{center}
  \includegraphics[width=15cm]{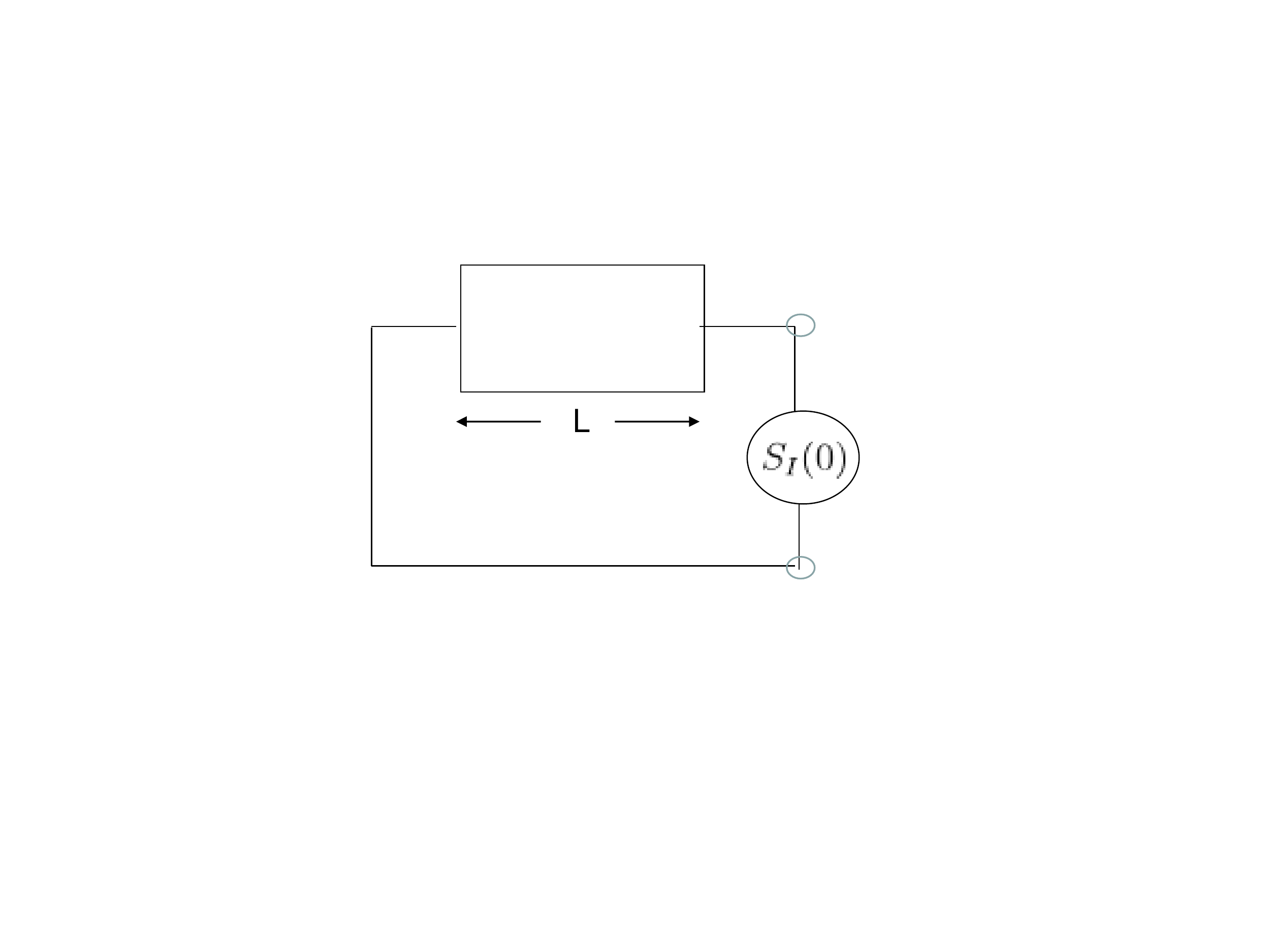}
 \end{center}
\vspace{-30mm}
 \caption{ Schematic of the circuit used to determine conductance  from current fluctuations due to fluctuations of the total carrier number transmitted through the sample, as measured in the external short-circuit of the open system.}
\label{figsi}
\end{figure}
From duality, following  Nyquist relations \cite{nyquist28}  
conductance is related to fluctuations of current measured on the external short circuit $\delta I(t)$, and resistance as fluctuations of voltage  measured on the external open circuit $\delta V(t)$ \cite{reggiani16} as schematically reported in Figs. (\ref{figsi}, \ref{figsv}).
\par
The expressions relating conductance and resistance  to fluctuations 
are fully compatible with the Landauer view that conductance is transmission, further adding the dual view that resistance is reflection. 
\par
Within the framework of the representation of conductance and resistance in terms of microscopic noise sources, we notice that the balistic regime is implicitly given in the derivation, since the concept of friction associated with a relaxation time is not needed to define conductance and resistance, here this concept is replaced by the stochastic characteristics of transmission and reflection, i.e. no fluctuations no response and the conversion of $\tau$ into a deterministic transit time for the balistic regime is fully legitimate.  
Thus, for a 1D geometry the balistic classical (bal,c) conductance/resistance are given by:
\begin{equation}
G^{bal,c} =1/R^{bal,c}= \frac{e^2 N} {Lm \sqrt{v_x^{'2}}} 
\label{GRb}
\end{equation}
with $N$ being he 1D carrier number.
\par
For completeness, Table 1 reports a brief historical overview of the electrical conductance/resistance concept starting from Volta discovery in 1799 of the first static electrical-energy generator that made possible Ohm experiments. 
\par
\begin{table}[pt]
\caption{Brief historical overview of the electrical conductance/resistance concept of a homogenous sample of volume $V=A L$.}
{\begin{tabular}{@{}ccccc@{}} \toprule \\
		Year & Author 
		\\ \colrule
		1799 & Volta & Invention of voltage dc supplier   \\
		1826 & Ohm & $V=RI$ law  \\
		1865 & Maxwell & Homogeneous dielectric impedance \\
		1900 & Drude &  Conductance kinetic model  \\
		1916 & Sommerfeld & fine structure constant $\alpha = 1/137$ \\
		1928 & Nyquist  & $I$, $V$  Noise spectral densities   \\
		1957 & Landauer & Conductance is transmission \\
		1980 & Von klitzing & High accuracy of quantum unit of conductance  \\
		2016 & Reggiani Alfinito Kuhn & Conductance/resistance from fluctuations of carrier number/drift-velocity  \\
		2018 & CODATA Value & Vacuum radiation impedance   \\
		2023 & Reggiani Alfinito Intini & Resistance is reflections \\
	\botrule	
		\end{tabular}}
\label{tab:BriefHistoricalOverview}
\end{table}
\par
We remark that within the Drude diffusion model  $G$ and  $R$ satisfy the reciprocity relation $GR=1$,
while within the FD model, they also exhibit  the duality property  that for the noise sources writes:
\begin{equation}
\overline{ \delta N^2}  \ \overline { v^{2'}_x}  =
\overline{ N}^2  \ \overline{ \delta v_d^2} \
= \frac{\overline{N} K_BT}{m}  
\label{N2v2}
\end{equation}
\par
Furthermore, in the limit of zero temperature   the duality property exhibited by conductance/resistance no longer holds, and we obtain:
\begin{equation}
lim_{T \rightarrow 0} \ G^{FB} =	G^{dif}
= \frac{e^2 \tau \overline {N} }{m L^2}
\end{equation}
that is, in the limit of zero temperature $ G^{FB} $ and $R^{FB}$ are given by the Drude formula so that, in the absence of thermal equilibrium conditions the duality property no longer holds.

\subsection{The one-dimensional balistic case, from classical 
to quantum conductnce} 
The one dimensional balistic case is obtained by setting in Eq.(\ref{GRb})
$\Gamma=1$ and,
under  quantum conditions  the $T=0$ limit  implies 
\begin{equation}
m \sqrt{\overline {v_x^{'2} }} L =h
\end{equation}
with $h$ the Planck constant.-
\par
Further, by considering energy quantization along the transverse direction, carrier number is substituted by the sum over the $i$-th transverse mode
as: 
\begin{equation}
{\overline N}=2
\end{equation}
thus obtaining the fundamental quantum unit of conductance $G_0$.
\subsection{The case of vacuum}
For completeness we recall that vacuum classical-electrodynamics gives
for the vacuum conductance $G_{vac}$:
\begin{equation}
G_{vac}=\frac{1}{R_{vac}}
 =\epsilon_0 c 
\label{Gv}
\end{equation}
Remarkably, the following quantum interrelation defines the fine structure constant $\alpha$ \cite{sommerfeld16}	:
\begin{equation}
\alpha = \frac{G_0}{4 G_{vac}} 
= \frac{e^2 }{2 h \epsilon_0 c } = \frac{1}{137}
\end{equation}
with $c$ the light velocity in vacuum.
\section{Conclusions and remarks} 
In this paper we compare the microscopic interpretation of Ohmic conductance and resistance of a two-terminals homogeneous conductor determined by the response to external macroscopic weak perturbations or by internal microscopic sources of thermal fluctuations described in terms of Nyquist relations \cite{reggiani16,nyquist28}.
Several models (classical, quantum, diffusive, ballistic, etc) are considered. 
On the one hand, making use of the thermal
noise sources for current fluctuations in an open system the Landauer \cite{landauer57} paradigm that conductance is associated
with transmission is clearly confirmed in he classical form, see Eq. (\ref{eGc}). 
On the other hand, looking at the noise sources for voltage fluctuations
in a closed system this leads to the dual concept that resistance is associated with reflection, see Eq. (\ref{eGc}). If transmission
refers to carriers randomly injected through  transparent contacts and transmitted to the opposite contact, then reflection
refers to carriers that are reflected at the contacts from inside the sample  or from outside the sample in the case of a dielectric medium, in particular the vacuum.
This extends the Landauer paradigm to the case of macroscopic conductors in the presence of a temperature with resistance viewed as reflection.
From a thermal equilibrium point of view, conductance follows from an open system,  i.e. a grand-canonical ensemble.
By contrast, resistance follows from a closed system, i.e. a canonical ensemble.
We remark that in both the cases conductance and resistance keep their definition also for balistic systems, that is in the absence of scattering mechanisms inside the sample. In other words, conductance and resistance are a consequence of carriers in their motion  between ideal contacts acting as ideally transparent to carrier number transmission for the case of conductance and ideally reflecting to carrier drift-velocity for the case of resistance.  The presence of scattering among carriers or with impurities inside the sample contributes to decrease the transit time thus leading to lower (to increase) the conductance (the resistance) with respect to the case of a balistic transport regime. 
These conclusions held in particular for the case of 1D quantum conditions, where the fundamental unit of conductance (or resistance) refer to  balistic conditions (i.e. absence of interactions).  Interactions inside the quantum system can be accounted for by including a transmission coefficient, or more 
generally a scattering matrix as developed originally within the so called Landauer-Buttiker formalism \cite{landauer57,buttiker86}.  
\par
Main points that received new insights in specific parts  of the paper are briefly summarized in the following list.
\par\noindent
1 - The duality and reciprocity relations between microscopic noise sources responsible of the so-called thermal agitation of electric charge in conductors have been investigated.
Here, fluctuations of the total number of carriers inside the physical system are shown to be responsible of current fluctuations detected in the external short-circuit, while fluctuations of the carrier drift-velocity are found to be responsible of the voltage fluctuations detected between contacts in the open external circuit as Johnson  noise \cite{johnson28}.
In essence, the duality relations imply  a generalized  Biot-Savart law converting the variance of current fluctuations with the variance of magnetic field fluctuations,  and a generalized Ohm law converting the variance of drift-velocity fluctuations with the variance of electric field fluctuations.
\par\noindent
2 - When moving from conductors to dielectrics or vacuum media the fundamental unit of conductance (resistance) reduce  to the definition of radiation (intrinsic) impedance  $\eta$ given by:
\begin{equation}
\eta= =\sqrt{\frac{\mu_0 \mu_r}{\epsilon_0 \epsilon_r}} 
\end{equation}
with $\mu_r$ and $\epsilon_r$, respectively  the relative permeability and permittivity of the medium,     
once the electrical charge is substituted by the vacuum (or Planck) charge making use of the fine-structure constant. 
\par\noindent
3 - Within this paper we defined conductance and is reciprocal resistance 
associated with several physical phenomena as: 

- linear response described by Ohm law 

- fluctuations phenomena at thermal equilibrium at a given temperature described by Nyquist relations

- diffusion phenomena described  by the generalized Einstein relation

- thermal phenmena described by Wiedemann-Franz law

- mesoscopic  quantum phenmena described by Landauer-Buettiker law

- Maxwell electromagnetic fields describing  vacuum impedance.
\section{Acknowledgments}
Prof. Tilmann Kuhn from M\"unster University is warmly thanked for the very valuable comments he provided on the  subject. 
%

%

%

\end{document}